\documentclass[3p,twocolumn,preprint, sort&compress]{elsarticle}

\journal{Physics Letters B}

\usepackage{hyperref}
\usepackage{graphicx}
\usepackage{subcaption}
\usepackage{amssymb}
\usepackage{amsmath}
\usepackage{yfonts}
\usepackage{comment}
\usepackage[dvipsnames]{xcolor}
\usepackage{multirow}
\usepackage{hhline}
\usepackage{booktabs}

\newcommand{\hc}[1]{#1^{\dagger}}
\newcommand{\abs}[1]{\lvert #1 \rvert}
\newcommand{\Neff}{N_{\rm eff}}
\newcommand{\mer}{m_{\rho} }
\newcommand{\meh}{m_{h} }
\newcommand{\er}{\rho }
\newcommand{\eh}{h }
\newcommand{\hu}{km s${}^{-1}$ Mpc${}^{-1}$ }

\newcommand{\gs}{g_{\star}}
\newcommand{\Tg}{T_{\gamma}}
\newcommand{\MP}{\bar{M}_{\rm Pl}}

\definecolor{KBFIred}{RGB}{163,35,47}
\newcommand{\lm}[1]{\textcolor{KBFIred}{\textbf{#1}}}
\date{\today}

\begin{document}

\begin{frontmatter}

\title{
\lm{A minimal model of inflation and dark radiation}}

\author[]{Kristjan Kannike\corref{mycorrespondingauthor}}
\cortext[mycorrespondingauthor]{Corresponding author}
\ead{kannike@cern.ch} 
\author[]{Aleksei Kubarski}
\author[]{Luca Marzola}
\author[]{Antonio Racioppi}
\address{NICPB, R\"avala 10, 10143 Tallinn, Estonia.}

\begin{abstract}
	We show that a minimal extension of the Standard Model including a new complex scalar field can explain inflation and the observed effective number of neutrinos. The real part of the singlet plays the r\^ole of the inflaton field, while the Goldstone boson emerging from the spontaneous symmetry breaking of a global $U(1)$ symmetry 
	contributes to dark radiation and increases the effective number of neutrino species by $0.3$ over the Standard Model value. After detailing the phenomenology of the model, we find that the predicted inflationary observables are in agreement with the current bounds, once the dark radiation component is allowed, both within the metric and Palatini formulation of non-minimally coupled gravity.   
\end{abstract}


\end{frontmatter}

\section{Introduction}
\label{sec:intro}

According to contemporary understanding, the earliest stage of our Universe presents a period of exponential expansion known as cosmic inflation. In the simplest models, the expansion is driven by a scalar field -- the inflaton -- that slowly rolls along its potential. At the end of inflation the inflaton decays, filling and (re)heating the Universe with matter and radiation~\cite{Starobinsky:1980te,Guth:1980zm,Linde:1981mu}, while its quantum fluctuations  seed the density perturbations that give rise to the observed cosmological structure~\cite{Olive:1989nu,Lyth:1998xn}.  

The evolution of these perturbations, however, depends on the temperature and  particle content of the primordial plasma: relativistic particles tend to escape from gravitational wells and consequently hinder the progressive formation of structures. Measurements of the characteristic scales of the Universe, in particular observations of the cosmic microwave background (CMB) and baryon acoustic oscillations (BAO), can therefore put a cap on the maximal number of such particles allowed at a given era. Taking the Standard Model (SM) as a reference, the possible presence of new relativistic degrees of freedom at a given temperature is usually quantified in the \emph{effective number  $\Neff$ of relativistic fermionic} -- or `neutrino' -- \emph{species}. 

Current CMB observations are certainly consistent with the SM prediction $\Neff^{\rm SM} = 3.046$, which considers only the three active neutrino flavours that participate in weak interactions.\footnote{Decimals are the result of a  non-instantaneous neutrino decoupling and quantum corrections.} The latest Planck results combined with BAO observation report in fact $\Neff =  2.99\pm 0.17$~\cite{Aghanim:2018eyx}, and seemingly leave little room for any \emph{dark radiation}: new relativistic free-streaming degrees of freedom that would add to the SM neutrino contribution to $\Neff$. 

The robustness of this conclusion has recently been questioned in connection to a standing anomaly concerning the measurements of the Hubble parameter $H_0$. Whereas CMB and BAO studies support $H_0 = 67.66 \pm 0.42$ \hu at a 68\% confidence level (CL) ~\cite{Aghanim:2018eyx}, local determinations of this quantity from supernova observations and the recently released GAIA data actually favour a larger value, $H_0 = 73.52 \pm 1.62$ \hu, giving rise to a 3.8 $\sigma$ discrepancy~\cite{Riess:2018byc,2018A&A...616A...1G}. Interestingly, this inconsistency could be diminished by allowing for dark radiation at the decoupling epoch, as the same CMB data prefer larger values of $H_0$ in this case. A combined fit of the Planck data accounting for the local $H_0$ determination results, for instance, in $H_0 = 69.32 \pm 0.97$ \hu for $\Neff=3.27 \pm 0.15$, both at a 68\% CL~\cite{Aghanim:2018eyx}.

If we entertain the idea that larger values of $\Neff$ are the solution to the Hubble parameter discrepancy, the corresponding change in the energy balance will modify the expansion history of the Universe. Big Bang nucleosynthesis (BBN), as an example, is sensitive to the presence of extra degrees of freedom at a time much earlier than radiation decoupling. The latest analyses report that reproducing the correct primordial abundances of light elements constrains $\Neff = 2.88\pm 0.27$ at a 68\% CL~\cite{Pitrou:2018cgg}, and therefore leave open the possibility that a new relativistic component solve the Hubble parameter puzzle.\footnote{We should nevertheless remark that additional properties such as viscosity and  speed of sound related to a possible new relativistic component are tightly constrained by the study of CMB perturbations~\cite{2012PhRvD..85b3001S}.}
Besides BBN, modifications of $\Neff$ -- thus of $H_0$ -- bear consequences also on the inflationary observables~\cite{Tram:2016rcw, Guo:2017qjt}. Here, dark radiation strongly affects model selection, as the damping of the CMB anisotropy power spectrum induced at small scales can be partially compensated by shifting the tilt of the primordial spectrum $n_{s}$ towards its blue side. For instance, once dark radiation is allowed, models of power-law inflation and curvaton scenarios can be reconciled with the present observation~\cite{Tram:2016rcw}.   

A Goldstone boson resulting from the breaking of a global symmetry is a natural candidate for the r\^{o}le of dark radiation~\cite{Weinberg:2013kea}. Its contribution to the effective number of neutrino species depends on the temperature at which the Goldstone leaves thermal equilibrium with the SM plasma, which is reheated by particle annihilations several times during the evolution of the Universe. Fascinated by this possibility, we study a minimal scenario where the dynamics of inflation are entwined with the properties of dark radiation. In our scheme,\footnote{An alternative scenario based on a  non-minimally coupled SM Higgs boson as the inflaton and $N$ Goldstone bosons is discussed in Ref.~\cite{Cheng:2018axr}.} a $U(1)$ global symmetry, which conserves the particle number of a complex scalar $S$, is broken as the radial part of the latter develops a vacuum expectation value (VEV). The phase of $S$ is the Goldstone boson that contributes to $N_{\rm eff}$, whereas the radial component plays the r\^{o}le of the inflaton. We also introduce a non-minimal coupling between $S$ and the curvature scalar $R$, which flattens the potential at large field values and generally leads our inflationary solutions towards those of the Starobinsky model~\cite{Starobinsky:1980te}. In this way we show that the proposed framework can fit the observed effective number of neutrino species and predict inflationary observables in a consistent way. For the sake of generality, we perform our analysis considering both the metric and Palatini formalisms \cite{Bauer:2008zj}, taking into account the relevant radiative corrections.

This Letter is organized as follows. In Section~\ref{sec:model} we introduce our model; the results concerning the effective number of neutrino species are given in Section~\ref{sec:goldstone}. Inflationary observables are calculated in Section~\ref{sec:inflation} and our conclusions are presented in Section~\ref{sec:conclusions}. We finally include the renormalisation group equations of the model in~\ref{sec:rge}.

\section{Model}
\label{sec:model}

In addition to the SM Higgs doublet $H$, our scalar sector comprises a complex field $S$ that transforms as a singlet under the SM gauge group, but carries a charge associated with a new global $U(1)$ symmetry. We also let $S$ be non-minimally coupled to gravity:  
\begin{equation}
  \mathcal{L} \supset -\left( \xi \hc{S} S + \frac{M^{2}}{2} \right) R + \abs{\partial H}^{2}+ \abs{\partial S}^{2} - V(H,S),
  \label{eq:L}
\end{equation}
where the scalar potential is given by
\begin{equation}
\begin{split}
  V &= \mu_{H}^{2} \abs{H}^{2} + \mu_{S}^{2} \abs{S}^{2} +  \lambda_{H} \abs{H}^{4} 
  \\
  &+ \lambda_{S} \abs{S}^{4} + \lambda_{HS} \abs{H}^{2} \abs{S}^{2}\,,
\end{split}
\label{eq:pot}
\end{equation}
and we take $\mu_{S}^{2},\,\mu_{H}^{2}<0$ but\footnote{Models with $ \lambda_{HS} <0$ are also viable, but require a more careful analysis since too large negative values of the portal coupling might prevent a successful EWSB.} $ \lambda_{HS} >0$.
We anticipate that the VEV developed by the radial component of $S$ is much smaller than the reduced Planck scale $\bar{M}_{\rm Pl}  = 2.4 \times 10^{18}$~GeV, so we set $M \simeq \bar{M}_{\rm Pl}$ in order to recover the usual strength of gravitational interactions.

In the unitary gauge, we parameterize
\begin{equation}
  H = \begin{pmatrix}
    0 \\ \frac{v_{h} + h'}{\sqrt{2}}
  \end{pmatrix},
  \qquad
  S = \frac{v_{\rho} + \rho'}{\sqrt{2}} e^{i 2 \alpha},
\end{equation}
where $\rho'$ is the radial part of the singlet and $\alpha$ is the Goldstone boson. 

The VEVs of the involved fields are computed via the minimization conditions
\begin{align}
  2 v_{h}^2 \lambda_{H} + v_{\rho}^2 \lambda_{HS} + 2 \mu_{H}^{2} &= 0\,, \label{eq:min:H}
  \\
  v_{h}^{2} \lambda_{HS} + 2 v_{\rho}^2 \lambda_{S} + 2 \mu_{S}^{2}&= 0\,, \label{eq:min:S}
\end{align}
where we neglect quantum corrections. The VEV of the Higgs boson is set to $v_{h} = 246.22$~GeV in accordance with measurements.

The portal coupling included $\lambda_{HS}$ in eq.~\eqref{eq:pot} mixes the CP-even scalar fields after the electroweak symmetry breaking, giving the mass matrix
\begin{equation}
	\operatorname{M}^2_{\text{CP+}} =
	\begin{pmatrix}
		2 \lambda_H v^2_H & \lambda_{HS} v_H v_{\rho} \\
		\lambda_{HS} v_H v_{\rho} & 2 \lambda_S v_{\rho}^2
	\end{pmatrix}\,,
\end{equation}
written here in the field basis $(h', \rho')$. The Goldstone boson remains massless as expected. The mixing between $h'$ and $\rho'$ is quantified in the angle
\begin{equation}
  \tan 2 \theta = \frac{ \lambda_{HS} \, v_{h} v_{\rho}}{ \lambda_H v^2_H - \lambda_S v_{\rho}^2}\,,
\end{equation}
which enters the definition of the mass eigenstates $\eh$ and $\er$:
\begin{equation}
	\begin{pmatrix}
		\eh \\ \er
	\end{pmatrix}
	=
	\begin{pmatrix}
		\cos\theta & \sin\theta \\ - \sin\theta & \cos\theta 
	\end{pmatrix}
	\begin{pmatrix}
		h' \\ \rho'
	\end{pmatrix}
	\,.
\end{equation}
 LHC measurements then impose $\meh = 125.09$~GeV \cite{Aad:2015zhl} and bound the mixing angle to $\sin \theta \lesssim 0.5$ for $\mer<\meh$~\cite{Khachatryan:2016vau,Robens:2016xkb}, which is the case we consider here.\footnote{Whereas solutions with $\mer > \meh$ indeed exist, we find that perturbativity of the theory and the constraint in eq. \eqref{eq:Gold} strongly disfavour the case. For more details, we refer the reader to the discussion following eq. \eqref{eq:Gold}.} Notice that the effect of the above mixing is negligible during the inflation: as $\lambda_{HS}>0$, the inflaton progressively drives the Higgs boson to vanishing field values~\cite{Lebedev:2012sy}, possibly solving, in this way, the SM vacuum stability problem.

\section{Goldstone bosons and the effective
 number of neutrino species}
\label{sec:goldstone}

The effective number of neutrino species quantifies extra-SM inputs into the radiation energy density of the Universe in units of the neutrino contribution. 
Taking the photon temperature $\Tg$ as a reference, we have  
\begin{equation}
\rho_{\rm rad } = \frac{\pi^2}{30} \gs \Tg^4\,,	
\end{equation}
where the \emph{effective number of relativistic degrees of freedom}
\begin{equation}
	\gs 
	= 
	\frac 78 \sum_{\textfrak{f}} g_\textfrak{f} \left(\frac{T_\textfrak{f}}{\Tg}\right)^4
	+
	\sum_{\textfrak{b}} g_\textfrak{b} \left(\frac{T_\textfrak{b}}{\Tg}\right)^4
\end{equation} 
includes all the fermionic, $\textfrak{f}$, and bosonic, $\textfrak{b}$, relativistic species. The first term in the above equation then accounts for the contribution of SM neutrinos, as well as of dark radiation
\begin{equation}
	\gs \supset \frac{7}{4} \Neff \left(\frac{T_\nu}{\Tg}\right)^4 \,.
\end{equation}
We can isolate the latter by defining $\Delta N_\nu = \Neff - \Neff^{\rm SM}$, where
\begin{equation}
	\Delta N_\nu = \frac 12 \sum_{\widetilde{\textfrak{f}}} g_{\widetilde{\textfrak{f}}} \left(\frac{T_{\widetilde{\textfrak{f}}}}{T_\nu}\right)^4
	+\frac 47
	\sum_{\widetilde{\textfrak{v}}} g_{\widetilde{\textfrak{b}}} \left(\frac{T_{\widetilde{\textfrak{b}}}}{T_\nu}\right)^4\,, 
\end{equation}
for the extra -- indicated by a tilde -- fermionic and bosonic species.

The presence of a Goldstone boson can thus result in different values of $\Neff$, depending on the temperature of these particles relative to that of SM neutrinos. Goldstone bosons that decouple from the SM plasma at the same time as the neutrinos yield $\Neff=3.617$, which is too large according to current observations. An earlier decoupling induces instead a smaller contribution, as the temperature ratio is suppressed by entropy injections in the plasma that reheat the neutrinos. For instance, Goldstone bosons that leave the thermal bath prior to the muon annihilation era result in $\Neff=3.438$~\cite{Weinberg:2013kea}, whereas a loss of thermal equilibrium after the QCD phase transition but preceding the pion annihilations yields $\Neff=3.350$. In this work we will consider the latter case, as earlier decouplings induce only negligible contributions to $\Neff$.   
 
To ensure that the Goldstone boson stays in thermal equilibrium until the required era, we need to take care that the rate of the thermalising interactions does not succumb to the expansion rate of the Universe. If we suppose that dark radiation be thermalised via effective interactions with the SM fermions -- mainly muons -- induced by the mixing of the CP odd scalar state, requiring that Goldstone decouple at a temperature of the order of the pion masses implies~\cite{Weinberg:2013kea}

\begin{equation}
  \frac{\lambda_{HS}^{2} m_{\mu}^2 m_\pi^5 M_{\rm Pl}}{m_{\rho}^{4} m_{h}^{4}} = \chi\,,
  \label{eq:Gold}
\end{equation}
where $ M_{\rm Pl}=\sqrt{8\pi} \bar{M}_{\rm Pl}$ is the Planck mass.
Here the factor $\chi$ corrects the above estimate for errors due to the considered simplified decoupling dynamics and possible mis-determinations of the effective number of degrees of freedom in the plasma~\cite{Ng:2014iqa,Olechowski:2018xxg,Garcia-Cely:2013nin,Drees:2015exa}. 
In this work we employ the above condition to set the value of the portal coupling $\lambda_{HS}$, and vary $\chi$ in the range $[0.1, \,10]$ to assess the effect of the mentioned uncertainties on the inflationary observables.

\section{Inflationary Observables}
\label{sec:inflation}

In order to delineate the inflationary scenario supported by the dynamics of $S$, we retain the quantum corrections\footnote{Whereas cosmological perturbations are invariant under frame transformations, see for instance Refs.~\cite{Prokopec:2013zya,Jarv:2016sow}, the equivalence of the Einstein and Jordan frames at the quantum level is still to be established. In the present letter we therefore adopt the following strategy: first we compute the effective potential in the Jordan frame, eq.~\eqref{eq:Veff}, and consequently we move to the Einstein frame for the calculation of the slow-roll parameters. Given a scalar potential in the Jordan frame, the cosmological perturbations are then independent, in the slow-roll approximation, of the choice of the frame in which the inflationary observables are computed~\cite{Prokopec:2013zya,Jarv:2016sow}. For a further discussion on frames equivalence and/or loop corrections in scalar-tensor theories we refer the reader to Refs.~\cite{Jarv:2014hma,Kuusk:2015dda,Kuusk:2016rso,Flanagan:2004bz,Catena:2006bd,Barvinsky:2008ia,DeSimone:2008ei,
Barvinsky:2009fy,Steinwachs:2011zs,Chiba:2013mha,George:2013iia,Postma:2014vaa,
Kamenshchik:2014waa,George:2015nza,Miao:2015oba,Inagaki:2015fva,Burns:2016ric,
Hamada:2016onh,Fumagalli:2016lls,Artymowski:2016dlz,Fumagalli:2016sof,Bezrukov:2017dyv,Karam:2017zno,Narain:2017mtu,Ruf:2017xon,Markkanen:2017tun,Ohta:2017trn,Ferreira:2018itt,Karam:2018squ}.}  in the inflaton potential and investigate its phenomenological consequences in both the metric and Palatini formalisms, which differ because of the presence of a non-minimal gravitational coupling \cite{Bauer:2008zj} as described in~\ref{sec:Palatini}.

Given the present constraint on the amplitude of scalar perturbations \cite{Ade:2013zuv,Ade:2015lrj} 
\begin{equation}
A_{s} = (2.14 \pm 0.05) \times 10^{-9}, \label{eq:As}
\end{equation} 
the running of the non-minimal coupling $\xi$ can instead be safely neglected in the computation as long as the pertubativity of the theory is ensured -- see for instance Ref.~\cite{Marzola:2016xgb}.

Within the metric formalism, the inflationary observables $r$ and $n_s$ can be computed by following straightforwardly the procedure delineated in Ref.~\cite{Jarv:2016sow}. The same procedure holds in the Palatini case, but the required field redefinition is here given by the solution of
\begin{equation}
	\frac{\partial \rho_E}{\partial \rho'} = \frac{\bar{M}_{\rm Pl}}{M} \sqrt{\frac{M^2}{M^2 + \xi (\rho')^2}} \, , 
	\label{eq:rhoE:Palatini}
\end{equation}
where $\rho_E$ is the canonically normalized Einstein-frame field value of the inflaton \cite{Bauer:2008zj} -- see \ref{sec:Palatini}. 

The total number of parameters that enter the computation amounts to the five coefficients in the scalar potential, the two VEVs of the CP even scalar fields, the $\chi$ parameter in eq.~\eqref{eq:Gold} and the non-minimal coupling to gravity, for a total of nine. The parameter space is however restricted by the two minimization conditions in eqs.~\eqref{eq:min:H} and~\eqref{eq:min:S}, the constraint in eq.~\eqref{eq:Gold}, the measurement of the Higgs mass and VEV and of the amplitude of scalar perturbations in eq.~\eqref{eq:As}, leaving only three free parameters. We choose these to be the inflaton VEV $v_{\rho}$, the mixing angle $\theta$ and the correction factor $\chi$.

Let us now discuss the implications of eq.~\eqref{eq:Gold} on the parameters space. As anticipated, we assume here that $\mer < \meh$, as in the complementary case $\mer > \meh$ such a constraint implies
\begin{equation}
  \lambda_{HS} > \frac{ m_h^4}{m_{\pi }^{5/2} m_{\mu } \sqrt{M_{\rm Pl}}} \sqrt{\chi} \simeq 92 \sqrt{\chi} ,
  \label{eq:portal:bound}
\end{equation}
corresponding to large values of the portal coupling that we choose to disregard for the sake of the perturbativity of the theory. 

By using eq.~\eqref{eq:Gold}, the inflaton mass can be approximated at the first order for a small mixing angle as
\begin{equation}
  \mer^2 \simeq \frac{m_{\pi }^{5/2} \sin \theta \; m_{\mu } \sqrt{M_{\rm Pl}}}{\sqrt{\chi } v_h v_{\rho}} ,
  \label{eq:inflaton:mass}
\end{equation} 
and the requirement that the inflaton itself do not contribute to dark radiation then imposes an upper bound on its VEV for given $\chi$ and $\sin\theta$. In our computation we conservatively require $\mer > 150$ MeV to prevent this possibility.

For the values of the inflaton mass selected by the constraints, the inflaton potential is dominated by the quartic coupling term, and the inflationary observables are consequently insensitive to the value of $\mer$. The parameters $\lambda_S$ and $\lambda_H$ can be derived by solving the minimization conditions in eq.~\eqref{eq:min:H} and~\eqref{eq:min:S} and by using the equations defining the mass eigenstates.

With the parameters of the potential being set at the electroweak scale, the  computation of inflationary observables cannot neglect the running of these quantities. Therefore, we write the Jordan frame inflaton potential as
\begin{equation}
V_{\rho} = \frac{1}{4} \lambda_S(\rho) \rho^4 \, , \label{eq:Veff}
\end{equation}
where $\lambda_S(\rho)$ is obtained from the renormalization group equations given in \ref{sec:rge}. Since generally $\lambda_{S} \ll \lambda_{SH}(m_t) \ll 1$, we find an approximated solution as  
\begin{equation}
\lambda_S(\rho) \simeq \lambda_{S}(m_t) + \beta_{\lambda_{S}} (m_t) \ln\left( \frac{\rho}{m_t}\right),
\end{equation}
where the parameters $\lambda_{S}(m_t)$ and $\beta_{\lambda_{S}}(m_t)$, for $m_\rho \ll m_h$, can be given as: 
\begin{align}
  \lambda_{S}(m_t) &\simeq \frac{v_h^2 \lambda _{H S}^2}{2 m_h^2} \simeq \frac{m_h^2 \sin ^2\theta}{2 v_{\rho}^2}, \label{eq:lambda:approx}
  \\
  \beta_{\lambda_{S}} (m_t) &\simeq \frac{\lambda_{HS}^{2}}{8\pi^2} \simeq\frac{m_h^4 \sin ^2\theta}{8 \pi ^2 v_h^2 v_{\rho}^2}. \label{eq:beta}
\end{align}

Furthermore, following Ref.~\cite{Marzola:2016xgb,Racioppi:2018zoy}, the running $\lambda_S$ can be approximated as
\begin{equation}
\lambda_S(\rho) \simeq \frac{m_h^2 \sin ^2\theta}{2 v_{\rho}^2} \left[1 + \delta \ln \left(\frac{\rho}{m_t}\right)\right],
\end{equation}
with 
\begin{equation}
\delta = \frac{m_h^2}{4 \pi ^2 v_h^2} \simeq 6.55 \times 10^{-3}.
\end{equation}
When the above approximations hold, we expect that the inflationary predictions in both the formulations of gravity lie on the corresponding $\delta \simeq 6.55 \times 10^{-3}$ lines, regardless of the values of $\chi$, $v_{\rho}$ and $\sin\theta$. The exact location of each point on a line is nevertheless determined by $v_{\rho}$ and $\sin\theta$ parameters, while the $\chi$ parameter generally affects the resulting inflaton mass only.\footnote{The correction factor $\chi$ modifies the inflationary observables only in the few cases where the running of $\lambda_{S}$ is dominated by its self-corrections, in which case the renormalization group equations need to be solved numerically.}

We can also estimate a rough lower bound for $v_{\rho}$ by imposing  $\lambda_{S}(m_t)<1$, a necessary condition for the theory to retain perturbativity up to the inflation scale. By using eq.~\eqref{eq:lambda:approx} we then obtain  
\begin{equation}
v_{\rho} > \frac{m_h}{\sqrt 2} \sin\theta \simeq 90 \sin\theta \text{ GeV} .
\end{equation}

In our computation we took $\sin\theta\in\{$0.001, 0.01, 0.1, 0.15, 0.2, 0.25, 0.3, 0.35, 0.4, 0.45, 0.5$\}$, whereas $v_{\rho} \in \{90, 10^2,10^3,10^4,10^5,10^6\}$ GeV. As mentioned before, we let the parameter $\chi$ in eq.~\eqref{eq:Gold} take values in the range $[0.1, \, 10]$ and we excluded points resulting in a inflaton contribution to dark radiation by imposing $\mer > 150$ MeV. 

\subsection*{Reheating and duration of inflation}
\label{sec:reheating}

After inflation, the inflaton oscillates around its VEV and consequently decays starting the so-called reheating process. In our scenario, the main decay mode of the inflaton is a pair of SM fermions with $m_f < m_\rho/2$, or a pair of Goldstone bosons, if $v_\rho$ is within $\mathcal{O}(10^{2} m_\rho)$. The former process is allowed by the mixing with the SM Higgs boson induced by the portal coupling.
Knowledge of the reheating dynamics allows an estimate of the duration of the inflationary era, which lasts for a number of $e$-folds approximately given by~\cite{Ellis:2015pla}
\begin{equation}
\begin{split}
  N_e &\simeq 61.1181 + \frac{1}{12} \frac{(1 - 3 \omega_{\text{int}} \ln 
  \frac{\rho_{\text{RH}}}{\rho_{\text{end}}})}{\omega_{\text{int}} + 1}
  \\
  &+ \frac{1}{4} \ln \frac{V_{*}^{2}}{\rho_{\text{end}} M_{\rm Pl}}.
\end{split}
\end{equation}
Here the $e$-fold average of the equation-of-state parameter during the thermalization epoch, $\omega_{\text{int}}$, and the energy density after reheating, $\rho_{\text{RH}}$, are respectively given by 
\begin{equation}
  \omega_{\text{int}} \simeq \frac{0.782}{ \ln \frac{2.096 m_{\rho}}{\Gamma_{\rho}}},
  \quad
  \rho_{\text{RH}} \simeq 0.0151 \Gamma_{\rho}^{2} M_{\rm Pl}^{2},
\end{equation}
where $\Gamma_\rho$ is the total decay width of the inflaton. The energy density $\rho_{\text{end}}$ at the end of inflation is instead computed by setting  $\epsilon_{H} =1$, which yields $\rho_{\text{end}} = \frac{3}{2} V_{\text{end}}$, where $V_{\text{end}}$ is the inflaton potential at the end of inflation.

\subsection*{Results}
\label{sec:results}

\begin{figure*}[htbp]
  \centering
          \includegraphics{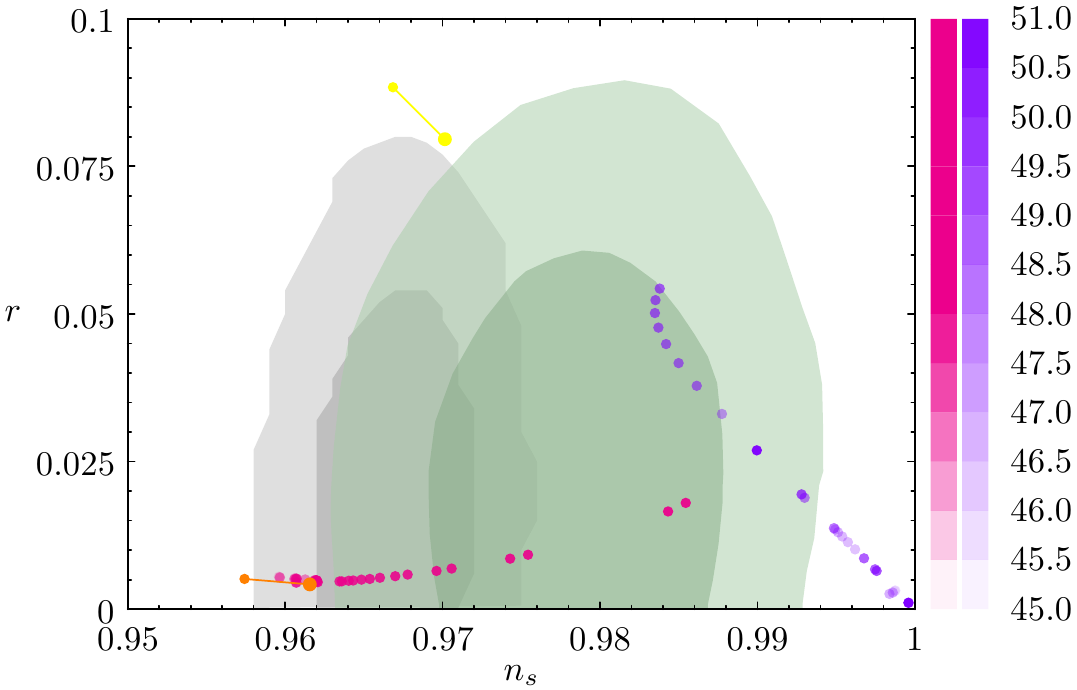}
        \caption{Inflationary observables $r$ and $n_s$. The predictions of the model are marked in magenta for the metric and in purple for the Palatini gravity, assuming $\chi=1$. In both cases, the color scale indicates the number of $e$-folds obtained. For the purpose of reference we show also the predictions of the Starobinsky $R^2$~\cite{Starobinsky:1980te} -- in orange --  and linear inflation~\cite{Akrami:2018odb} -- in yellow -- models, with smaller dots corresponding to a duration of inflation of 45 $e$-folds and larger  dots to 50 $e$-folds. The grey areas indicate the 68\% and 95\% CL supported by the Planck 2018 data~\cite{Akrami:2018odb} for a standard value of $\Neff$. Correspondingly, the green areas show how the same bounds shift once a dark radiation component -- $\Neff=3.3$ -- is allowed~\cite{Guo:2017qjt}.  }
      \label{fig:resuts}
    \end{figure*}

 \begin{table*}
    \begin{center}
    \begin{tabular}{c|c|c|c|c|c|c|c|c|c|c||c|c}
    \toprule
    \multicolumn{5}{c}{} &  \multicolumn{3}{|c}{metric} & \multicolumn{3}{|c|}{Palatini}\\
    \midrule
     $ v_\rho/\text{GeV} $  & $ \sin\theta  $          &  $\chi    $&   $ m_{\rho }/\text{GeV} $ & $ \lambda _S $  &$N_e$& $ n_s $ & $ r $ 				 & $N_e$&  $ n_s $ & $ r $  \\ \hline
    \multirow{3}{*}{$ 90 $} &\multirow{3}{*}{$ 0.4 $} &   $ 0.1 $  &   $ 11.7$ & $ 0.162 $  					  & $  51 $ & $ 0.986 $ & $1.89\times 10^{-2}$&  \multirow{3}{*}{$\star$}  & \multirow{3}{*}{$\star$}  & \multirow{3}{*}{$\star$}\\     
    & & $ 1   $                                                    &   $ 6.63 $ & $ 0.157 $ 					  & $ 50.6$ & $0.984 $ & $1.65\times 10^{-2}$&    &   &\\                                      
    & & $ 10  $                                                    &   $ 3.73 $ & $ 0.155 $ 					  & $ 50.3$ & $0.984 $ & $1.6\times 10^{-2} $&    &   &\\                                      
    \hline                                   
    \hline                                                                      
    \multirow{3}{*}{$ 10^3 $} &\multirow{3}{*}{$ 0.15 $}  & 0.1  & $ 2.25 $ & $ 1.79\times 10^{-4} $ & $ 48.7 $&$ 0.962 $ & $4.83\times 10^{-3}$ & $ 48.7 $ & $0.988  $&$ 3.28\times 10^{-2} $\\ 
     & & 											1     & $ 1.27 $ & $ 1.77\times 10^{-4} $  & $ 48.4 $&$ 0.962 $ &$ 4.9\times  10^{-3} $& $       48.4 $& $0.988  $&$ 3.31\times  10^{-2} $\\
     & & 											10    & $ 0.71 $ & $ 1.76\times 10^{-4} $  & $ 48. $&$ 0.962  $&$  4.95\times 10^{-3} $&$        48.1$ &$ 0.988  $& $ 3.32\times 10^{-2} $\\
    \hline                                   
    \hline                                                                     
    \multirow{3}{*}{$ 10^5 $} &\multirow{3}{*}{$ 0.25 $}  & $ 0.1 $ & $ 2.88  $ & $ 4.89\times 10^{-8} $ & $ 46.1 $ & $ 0.96  $ & $  5.43\times 10^{-3}    $ &  $ 45.5$ & $ 0.998  $ & $    2.58\times 10^{-3} $\\ 
& & 1 & $ 1.62  $ & $ 4.89\times 10^{-8} $ & $ 45.8 $ & $ 0.96  $ & $  5.49\times 10^{-3}    $ & $ 45.2 $ & $ 0.998  $ & $   2.59\times 10^{-3} $ \\ 
  & & 10 & $ 0.09  $ & $ 4.89\times 10^{-8} $ & $ 45.5 $ & $ 0.959  $ & $  5.56\times 10^{-3}   $  & $ 44.9 $ & $ 0.998 $ & $    2.59\times 10^{-3} $ \\
 \hline
    \end{tabular}
    \end{center}
    \caption{Benchmark points for the considered model. In correspondence of the first entry, the running of the inflaton quartic coupling is dominated by self-corrections. We indicated with $\star$ the appearance of a Landau pole below the scale of inflation. Notice that the points corresponding to the last entry of the table are in tension with the inflationary observables once dark radiation is assumed. The last point is also to be rejected, as in this case the inflaton contributes to dark radiation.   }
    \label{table}
    \end{table*}
 
Our results are presented in Figure~\ref{fig:resuts}, where the predicted values of $n_{s}$ and $r$ assume a value of the correction parameter $\chi=1$. In the following discussion we argue that, at the current accuracy, inflationary observables are basically independent of the choice of $\chi$. 

Within the metric formulation -- points in magenta --, as long as the quartic coupling self-corrections in the inflaton potential play a marginal r\^ole, the solutions of the model given by Eqs.~\eqref{eq:lambda:approx} and~\eqref{eq:beta} tend to the ones supported by the Starobinsky $R^2$ inflation -- shown in orange -- and are generally insensitive to the value assumed by the correction coefficient $\chi$ of eq.~\eqref{eq:Gold}. Explicit benchmark points exhibiting this behaviour are shown in the last two entries of Table~\ref{table}. 

In the opposite regime, when the running of $\lambda_S(m_t)$ is dominated by self-corrections, our predictions are drastically departing from the Starobinsky attractor and reach well within the 68\% confidence level (CL) region. In this case, inflationary observables may mildly depend
on the correction coefficient $\chi$ as illustrated by the first entry in Table~\ref{table}.
Therefore, within the metric formalism, we conclude that  the predictions of the model fall below the 95\% CL indicated by the data -- indicated by the light green area, assuming $\Neff=3.3$ -- 
when the running of $\lambda_S(m_t)$ is dominated by the same parameter. 

As for the case of Palatini formulation -- purple points in Figure~\ref{fig:resuts} -- the predictions of the model still select a small tensor-to-scalar ratio $r$, but support values of the spectral index $n_s$ generally shifted towards the blue end, which even overshoot the 95\% CL. We also see that the solutions approach the linear inflation attractor, in agreement with Ref.~\cite{Racioppi:2018zoy}, and that fewer solutions depart from the $\delta \simeq 6.55 \times 10^{-3}$ line with respect to the metric case. This is because the larger inflaton field excursion typical of the Palatini formulation results more easily in a Landau pole once the running of the inflaton quartic coupling is dominated by self-corrections -- see for instance the first entry in Table~\ref{table}. Moreover, we remark that the inflationary solutions within the Palatini formalism of the model with $\mer > 150$ MeV are in agreement with observations only if a dark radiation component is allowed, falling on the outside of the 95\% CL contour corresponding to a standard value of $\Neff$ -- shown by the light gray area.

To conclude, notice that both formalisms predict a number of $e$-folds relatively low, as expected for a light inflaton with a small decay width. The negligible difference in the number of $e$-folds computed in the two cases is explained by the low field values involved in the reheating dynamics, which are therefore rather insensitive to the performed field redefinition.

\section{Conclusions}
\label{sec:conclusions}

We proposed a model for inflation and dark radiation based on a minimal extension of the SM containing a complex scalar singlet. The scalar potential is symmetric under a global $U(1)$ symmetry, which is spontaneously broken by the dynamics of the new field. As a result, the modulus of the complex scalar field plays the r\^ole of the inflaton, whereas its phase yields a massless Goldstone boson which contributes to dark radiation. 

Considering a scenario where the Goldstone bosons leave the thermal bath immediately prior to the pion annihilation era, we computed the resulting effective number of neutrino species and detail the inflationary phenomenology predicted by the model for both the metric and Palatini formalism. 

For the considered effective number of neutrino species $\Neff = 3.3$, as long as the inflaton quartic self-corrections are marginal, we find that in the metric approach the predictions of the model tend to the Starobinsky limit regardless of details concerning the Goldstone decoupling dynamics. In the Palatini formalism, instead, the model generally selects values of the spectral index shifted toward the blue end of the spectrum. Small values of the tensor-to-scalar ratio are still favoured. The bulk of solutions for the Palatini formalism falls within the present 95\% confidence interval indicated by the data for a non-standard value of the effective number of neutrino species. When the self-corrections of the inflaton quartic coupling dominate the running of the parameter, the predictions of the metric case drastically depart from the Starobinsky attractor, reaching the core of the 68\% confidence region, and mildly depend on details in the dynamics of Goldstones decoupling. In this case, the Palatini formulation is instead generally ruled out by the appearance of Landau poles at scales lower than the inflationary one.
 
\subsection*{Acknowledgments}
This work was supported by the Estonian Research Council grant PUT799, PUT1026, the grant IUT23-6 of the Estonian Ministry of Education and Research, and by the EU through the ERDF CoE program project TK133.


\appendix

\section{Beta functions}
\label{sec:rge}
The running of interaction couplings is determined by the renormalisation group equations $\mu \frac{\mathrm{d} g_{i}}{\mathrm{d} \mu} = \beta_{g_{i}}$. 
We calculated the $\beta$-functions with the PyR@TE package \cite{Lyonnet:2013dna,Lyonnet:2016xiz} at one-loop level:
\begin{align}
  (4\pi)^{2} \beta_{g_{Y}} &=	\frac{41}{6} g_{Y}^{3},
  \\
  (4\pi)^{2} \beta_{g} &=	-\frac{19}{6} g^{3},
  \\
  (4\pi)^{2} \beta_{g_{3}} &=	-7 g_{3}^{3},
  \\
  (4\pi)^{2} \beta_{y_{t}} &= y_{t} \left( \frac{9}{2} y_{t}^{2} - \frac{17}{12} g_{Y}^{2} -\frac{9}{4} g^{2} 
  - 8 g_{3}^{2} \right),
  \\
  (4\pi)^{2} \beta_{\lambda_{H}} &=	\frac{3}{8} (3 g^{4} + 2 g^{2} g_{Y}^{2} + 3 g_{Y}^{4}) 
  - 6 y_{t}^{4}
  \notag
  \\
  &- (9 g^{2} -  3 g_{Y}^{2} + 12 y_{t}^{2}) \lambda_{H} 
  + 24 \lambda_{H}^{2} 
  \notag
  \\
  &+ \lambda_{HS}^{2},
  \\
  (4\pi)^{2} \beta_{\lambda_{HS}} &= \lambda_{HS} {\Big(} 4 \lambda_{HS} -\frac{9}{2}g^{2} 
  - \frac{3}{2} g_{Y}^{2}
  \notag
  \\
  & + 6 y_{t}^{2}  + 12 \lambda_{H} + 8 \lambda_{S} {\Big )},
  \\
  (4\pi)^{2} \beta_{\lambda_S} &=	20 \lambda_S^{2} + 2 \lambda_{HS}^{2}.
  \label{eq:beta:lambda:S}
\end{align}

\section{Palatini formalism}
\label{sec:Palatini}
The Palatini formalism of non-minimally coupled gravity follows from the Lagrangian 
\begin{equation}
\sqrt{- g} \mathcal{L} \!= \! \sqrt{-g}  \left(\!-\frac{\MP^2}{2}f(\rho')R(\Gamma) + \frac{(\partial \rho')^2}{2} \! - V(\rho') \! \right) ,
\label{eq:JframeL}
\end{equation}
where the connection $\Gamma$ is, a-priori, independent of the metric. When solved for a torsion-free connection, $\Gamma^\lambda_{\alpha\beta}=\Gamma^\lambda_{\beta\alpha}$, the associated equation of motion leads to~\cite{Bauer:2008zj}
\begin{equation}
 \Gamma^{\lambda}_{\alpha \beta} = \overline{\Gamma}^{\lambda}_{\alpha \beta}
+ \delta^{\lambda}_{\alpha} \partial_{\beta} \omega(\rho') +
\delta^{\lambda}_{\beta} \partial_{\alpha} \omega(\rho') - g_{\alpha \beta} \partial^{\lambda}  \omega(\rho') ,
\label{eq:conn:J}
\end{equation}
where 
\begin{equation}
  \label{omega}
  \omega\left(\rho'\right)=\ln\sqrt{f(\rho')},
\end{equation}
and $\bar{\Gamma}=\bar{\Gamma}(g^{\mu\nu})$ is the Levi-Civita connection, fully determined by the metric tensor through 
\begin{equation}
\label{eq:LC}
 \overline{\Gamma}^{\lambda}_{\alpha \beta} = \frac{1}{2} g^{\lambda \gamma} \left( \partial_{\alpha} g_{\beta \gamma}
+ \partial_{\beta} g_{\gamma \alpha} - \partial_{\gamma} g_{\alpha \beta}\right) .
\end{equation}

As general Lagrangians result in difference between the connection in eq.~\eqref{eq:LC} and \eqref{eq:conn:J}, the predictions of metric and Palatini formulations are usually distinguishable. One famous counter-example is given by the Einstein-Hilbert Lagrangian, in which case the Palatini formulation reduces to the metric one as the equation of motion of the connection forces $\Gamma^\lambda_{\alpha\beta} =\bar\Gamma^\lambda_{\alpha\beta}$.

The difference between the two formalisms persist also after the starting Lagrangian in eq.~\eqref{eq:JframeL} has been cast in the Einstein frame by means of the conformal transformation
\begin{eqnarray}
\label{eq:gE}
g_{\mu \nu}^E = f(\rho') \ g_{\mu \nu} .
\end{eqnarray}
When described in the Einstein frame, the gravity sectors of both the formalisms look alike (cf. eq. (\ref{eq:conn:J})), however the matter sector, consisting in our case of $\rho'$, behave differently. In more detail, the Einstein frame Lagrangian is given by~\cite{Bauer:2008zj}
\begin{equation}
  \sqrt{- g^{E}} \mathcal{L}^{E} = 
  \sqrt{- g^{E}} \Bigg[  - \frac{\MP^2}{2} R  + 
  \frac{(\partial \rho_E)^{2}}{2} - U(\rho_E) \Bigg] ,
   \label{eq:Einstein:Lagrangian}
\end{equation}
where $\rho_E$ is canonically normalized scalar field in the Einstein frame, with scalar potential
\begin{equation}
U(\rho_E) = \frac{V(\rho'(\rho_E))}{f^{2}(\rho'(\rho_E))} .
\label{eq:U}
\end{equation}

For the case of metric formulation, $\rho_E$ is derived by integrating the following relation
\begin{equation}
\frac{\partial \rho_E}{\partial \rho'} = \sqrt{\frac{3}{2}\left(\frac{\MP}{f}\frac{\partial f}{\partial \rho'}\right)^2+\frac{1}{f}} ,  
  \label{eq:dphiE}
\end{equation}
where the first term comes from the transformation of the Jordan frame Ricci scalar and the second one is due to the rescaling of the Jordan frame scalar field kinetic term. 

Differently, in the Palatini formulation, the field redefinition is induced only by the rescaling of the inflaton kinetic term, yielding 
\begin{equation}
\frac{\partial \rho_E}{\partial \rho'} = \sqrt{\frac{1}{f}} . 
  \label{eq:dphiP}
\end{equation}

As we can see, within the Einstein frame, the difference between the two formulations then amounts to the definition of $\rho_E$, which is reminiscent of the different non-minimal kinetic terms for $\rho'$ derived in these approaches.

%

\bibliographystyle{elsarticle-num}
\bibliography{neffinfl}

\end{document}